%% file: main.tex
\documentclass[runningheads]{llncs}

\input{packages}

\input{definitions}

\makeatletter
\newcommand{\printfnsymbol}[1]{%
  \textsuperscript{\@fnsymbol{#1}}%
}
\makeatother

\def\githuburl{\href{https://github.com/Janhutter/BSARec}{\texttt{https://github.com/Janhutter/BSARec}}}

\begin{document}
\title{A Systematic Reproducibility Study of BSARec for Sequential Recommendation}
\titlerunning{A Systematic Reproducibility Study of BSARec} 
\author{
Jan Hutter\thanks{Equal contribution; order was determined randomly.}\orcidlink{0009-0001-5833-8034} \and
Hua Chang Bakker\printfnsymbol{1}\orcidlink{0009-0008-0804-4426} \and
Stan Fris\printfnsymbol{1}\orcidlink{0009-0000-9300-5502} \and 
Madelon Bernardy\printfnsymbol{1}\orcidlink{0009-0007-0448-1816}\and
Yuanna Liu\orcidlink{0000-0002-9868-6578}
}

\authorrunning{Hutter et al.} 
\institute{
University of Amsterdam, The Netherlands\\
\email{\{j.w.j.hutter, h.c.bakker, s.c.j.fris, a.m.bernardy, y.liu8\}@uva.nl}
}

\maketitle

\input{sections/body}







\bibliographystyle{splncs04nat}
\bibliography{references}


\end{document}

%% file: packages.tex
\usepackage[T1]{fontenc}
\usepackage{graphicx}
\usepackage{color}

\usepackage{tabularx}
\usepackage{booktabs}
\usepackage{multirow}
\usepackage{adjustbox}
\usepackage{graphicx}
\usepackage{tikz}
\usepackage[inline]{enumitem}
\usepackage{acronym}
\usepackage{wrapfig}
\usepackage{capt-of}
\usepackage{amsfonts, amsmath, amssymb} 
\usepackage{mathtools}
\usepackage[numbers,sort&compress]{natbib}
\usepackage{caption}
\usepackage{float}
\usepackage{subcaption}

\usepackage{orcidlink}
\usepackage{hyperref}

\usepackage[capitalise]{cleveref}
\crefname{figure}{Figure}{Figures}
\crefname{enumi}{}{} 

%% file: definitions.tex
\newcommand{\header}[1]{\vspace*{2mm}\noindent\textbf{#1.}}

\acrodef{DFT}{discrete Fourier transform}
\acrodef{DWT}{discrete wavelet transform}
\acrodef{DC}{direct current}
\acrodef{SR}{Sequential recommendation}
\acrodef{BSA}{Beyond Self-Attention}
\acrodef{DSP}{digital signal processing}
\acrodef{NDCG}{normalized discounted cumulative gain}
\acrodef{HR}{hit rate}
\acrodef{SOTA}{state-of-the-art}

\acrodef{CNN}{convolutional neural networks}
\acrodef{RNN}{recurrent neural networks} 
\acrodef{MLP}{multilayer perceptron}

\newcommand{\negskip}{\vspace*{-1.5mm}}

%% file: sections/body.tex
\begin{abstract}
In sequential recommendation (SR), the self-attention mechanism of Transformer-based models acts as a low-pass filter, limiting their ability to capture high-frequency signals that reflect short-term user interests.
To overcome this, BSARec augments the Transformer encoder with a frequency layer that rescales high-frequency components using the Fourier transform.
However, the overall effectiveness of BSARec and the roles of its individual components have yet to be systematically validated.
We reproduce BSARec and show that it outperforms other SR methods on some datasets.
To empirically assess whether BSARec improves performance on high-frequency signals, we propose a metric to quantify user history frequency and evaluate SR methods across different user groups.
We compare digital signal processing (DSP) techniques and find that the discrete wavelet transform (DWT) offer only slight improvements over Fourier transforms, and DSP methods provide no clear advantage over simple residual connections.
Finally, we explore padding strategies and find that non-constant padding significantly improves recommendation performance, whereas constant padding hinders the frequency rescaler’s ability to capture high-frequency signals.

\end{abstract}
\section{Introduction}
Recommender systems effectively help users access content tailored to their preferences and enable platforms to broadly disseminate information~\cite{DBLP:journals/tois/FangZSG20,DBLP:conf/kdd/YingHCEHL18, DBLP:conf/sigir/0001DWLZ020}.
\ac{SR} aims to capture temporal dependencies in users’ historical interactions to predict the next item they are likely to engage with~\cite{DBLP:conf/cikm/HidasiK18, DBLP:conf/cikm/SunLWPLOJ19, DBLP:conf/icde/XieSLWGZDC22, DBLP:conf/icdm/KangM18, DBLP:conf/sigir/ChangGZHNSJ021}. 
Due to the dynamic nature of user preferences, \ac{SR} systems are expected to balance predictions between short-term interests within a session and long-term interests that develop across multiple sessions~\cite{DBLP:journals/is/BokaNN24}.

In \ac{SR}, deep learning-based methods such as \ac{RNN} \cite{DBLP:journals/corr/HidasiKBT15} and attention-based architectures~\cite{DBLP:conf/icdm/KangM18, DBLP:conf/cikm/SunLWPLOJ19, DBLP:conf/wsdm/QiuHYW22, DBLP:conf/sigir/DuYZQZ0LS23, DBLP:conf/aaai/Shin0WP24, DBLP:journals/ipm/SuCL25, yin2014icsrec} have been extensively investigated in recent years. Among them, self-attention has proven effective in capturing user preferences over time, and Transformer-based architectures~\cite{DBLP:conf/nips/VaswaniSPUJGKP17} have achieved \ac{SOTA} performance on \ac{SR} tasks~\cite{DBLP:conf/icdm/KangM18, DBLP:conf/aaai/Shin0WP24, DBLP:journals/ipm/SuCL25}.
Despite this success, a frequency-domain perspective has recently been introduced to provide new insights into user interest modeling~\cite{DBLP:journals/ipm/SuCL25}.
In the Fourier domain, long-term user interests can be interpreted as low-frequency signal in the interaction history, while short-term interests correspond to high-frequency components that fluctuate more rapidly~\cite{DBLP:conf/aaai/Shin0WP24}.
However, self-attention acts as a low-pass filter~\cite{DBLP:conf/iclr/WangZCW22, DBLP:conf/aaai/Shin0WP24}, which limits its ability to capture short-term interests.
To address this limitation, \citet{DBLP:conf/aaai/Shin0WP24} proposed BSARec, which incorporates a frequency-based inductive bias module via the Fourier transform to better capture high-frequency signal components, achieving \ac{SOTA} performance.
While BSARec offers a general solution to the low-pass filtering nature of self-attention, its effectiveness in modeling short-term user interests has not yet been empirically replicated. 
Moreover, the application of inductive bias module within attention-based recommendation frameworks remains to be fully explored.
Therefore, we rigorously reproduce BSARec and compare it with other \ac{SR} methods (\cref{rq:reproduction}), further extending the original study to provide deeper insights into its effectiveness.

\citet{DBLP:conf/aaai/Shin0WP24} claim that BSARec's ability to capture high-frequency signals in the history sequence. This should also be reflected in its performance when predicting for users with more short-term interests. To verify this hypothesis, we propose the \textbf{scaled DC component} metric to measure the frequencies of item categories in user historical interactions, and examine the performance of BSARec from the aspects of both user history frequency and target items (\cref{rq:history_frequency}).
We find that BSARec outperforms other \ac{SR} methods on high-frequency group in the LastFM and ML-1M datasets, noting that performance for different user groups varies strongly across different datasets.

Recent work~\cite{DBLP:conf/icsim/JiaDMQ25, mamba_wavelet, DBLP:conf/icic/LuGZZLG25, 10.1145/3731120.3744621} uses the \ac{DWT} for processing long-range dependencies, overcoming the Fourier transform’s limitation of operating solely in the frequency domain~\citep{boggess-2009}.
We further investigate this line of work (\cref{rq:dsp}) and include a comparison with a simple residual connection~\cite{DBLP:conf/cvpr/HeZRS16}.
We find that the choice of \ac{DSP} methods has only a limited impact on recommendation performance, and that comparable results can be achieved even when the \ac{DSP} method is omitted.

Zero-padding is necessary for the efficient processing of user histories~\cite{DBLP:conf/www/ZhouYZW22, DBLP:conf/sigir/DuYZQZ0LS23, DBLP:conf/aaai/Shin0WP24, DBLP:journals/ipm/SuCL25}, but it introduces low-frequency signals into the interaction sequence.
Motivated by this, we investigate the effect of different padding strategies on the performance of BSARec (\cref{rq:padding}) and find that alternative padding methods can provide significant performance improvements.

In this paper, we systematically verify and extend the analysis of~\cite{DBLP:conf/aaai/Shin0WP24} by investigating the following research questions:
\begin{enumerate}[label=\textbf{RQ\arabic*:},leftmargin=*, align=left,labelwidth=3em, labelsep=0.5em, ref={\textbf{RQ\arabic*}}]
    \item \label{rq:reproduction} Are the SOTA results of BSARec reproducible?
    \item \label{rq:history_frequency} To what extent do long- and short-term user interests impact the performance of SR models?
    \item \label{rq:dsp} What is the impact on BSARec performance of using \ac{DSP} methods other than the Fourier transform? 
    \item \label{rq:padding} To what extent can user history padding affect the performance of BSARec?
\end{enumerate}

\section{Related Work}

\subsubsection{\ac{DSP} for Sequential Recommendation.}
User histories often exhibit recurring interests in specific item categories (e.g., technological gadgets), while also encompassing abrupt and infrequent interests (e.g., home appliances).
Recurring interests are low-frequency signals in Fourier domain, while short-term interests inform the higher frequencies.
However, attention-based models have limited ability to detect high-frequency signals in user history, since attention acts as a low-pass filter~\cite{DBLP:conf/iclr/WangZCW22, DBLP:conf/aaai/Shin0WP24}.
Several methods~\cite{DBLP:conf/sigir/DuYZQZ0LS23, DBLP:conf/aaai/Shin0WP24, DBLP:journals/ipm/SuCL25, DBLP:conf/cikm/HidasiK18, DBLP:conf/sigir/WangLHYZZG25} have been proposed to address this issue using the Fourier transform.
FICLRec \cite{DBLP:journals/ipm/SuCL25} first transforms the low- and high-frequency components of the input using a \ac{MLP} before passing it to the attention layer.
BSARec \cite{DBLP:conf/aaai/Shin0WP24} provides an inductive bias module via a frequency rescaler, implemented as a parallel branch within the attention module, which rescales the high-frequency components of the input via the \ac{DFT}. 
As the \ac{DFT} discards time-domain information, several works propose to use the \ac{DWT} instead~\cite{DBLP:conf/icsim/JiaDMQ25, mamba_wavelet, DBLP:conf/icic/LuGZZLG25, 10.1145/3731120.3744621}.
An alternative approach to \ac{SR} relies solely on \ac{DSP} methods, \textit{without} the use of attention layers~\cite{DBLP:conf/www/ZhouYZW22, DBLP:conf/icic/LuGZZLG25, 10.1145/3731120.3744621, DBLP:journals/corr/abs-2312-00752, mamba_wavelet}.
Our work builds on the architecture of BSARec and investigates the effectiveness of different \ac{DSP} methods as a complement to self-attention.

\negskip
\subsubsection{Padding.}
Efficient batch processing of user histories requires padding the shorter sequences to match the pre-defined input sequence length.
In computer vision, the choice of padding strategy has been shown to significantly affect model performance~\cite{DBLP:conf/iciap/TangOB19, DBLP:journals/corr/abs-1903-07288}.
The effect of padding on user history representation in SR is less well-studied: previous research on applying signal analysis to \ac{SR} has employed either zero-padding~\cite{DBLP:conf/icic/LuGZZLG25, DBLP:journals/ipm/SuCL25, DBLP:conf/sigir/DuYZQZ0LS23, DBLP:journals/ijon/XuFZZWLS21, DBLP:conf/www/TanjimSBHHM20}, a special padding token~\cite{DBLP:conf/sigir/ParkKYHYCCC24} or a repeated version of the input sequence \cite{DBLP:conf/recsys/DangLYGJ0Z24}.
Any form of constant padding can be considered as low-frequency signals for \ac{SR} models, potentially influencing how the model processes the input, but this topic is yet to be explored for \ac{DSP}-based methods.
Our reproducibility study aims to address this gap.

\section{Preliminaries}\label{sec:preliminaries}
We first briefly review the \ac{DFT} and the \ac{DWT}.
The \ac{DFT} is a unitary transformation \(\mathcal{F} \colon \mathbb R^d \to \mathbb C^d\) that maps a sampled signal \(x \in \mathbb R^d\) to its representation \(\mathcal{F}(x) \in \mathbb C^d\) in Fourier domain, and can be represented by a matrix with the Fourier basis as rows; each corresponding to a different frequency.
For example, the \ac{DC} component \(\mathcal{F}(x)_1 = \sum_i x_i / \sqrt{d}\) represents zero frequency and can be interpreted as an average of the sampled signal.
Following \cite{DBLP:conf/aaai/Shin0WP24}, we consider the low-frequency components \(\operatorname{LFC}_c(x)\) to be the first \(c\) components of \(\mathcal{F}(x)\) with padded zeros to match the input dimension.
The high-frequency component of \(x\) is 
\begin{align}
    \operatorname{HFC}_c(x) \coloneqq x - \operatorname{LFC}_c(x).
\end{align}
The \ac{DFT} only allows for signal decomposition in the frequency domain, but not in the time domain~\cite{boggess-2009}. 
A different basis that retains temporal information can be parametrized by a wavelet function \(\psi\) which oscillates on a smaller subset of the input domain and is zero everywhere else.
The \ac{DWT} is similar to the \ac{DFT}, but uses the basis induced by a chosen wavelet function.
An example is the Haar wavelet \(\psi\), which is 1 if \(0 \leq t < 1/2\), -1 if \(1/2 \leq t < 1\) and 0 otherwise.

\negskip
\section{Methodology}
\negskip
\subsection{Problem Formulation}
\negskip
Let \(\mathcal{U}\) be the set of users and let \(\mathcal{I}\) be the set of items.
The history \(H_u = (h_1^u, \ldots, h_{n(u)}^u)\) of a user \(u \in \mathcal{U}\) is a chronologically ordered list of \(n(u)\) items \(h_i^u \in \mathcal{I}\) a user interacted with.
The length of the history \(n(u)\) is dependent on \(u\), but we omit this dependency for clarity and write \(n\) instead.
Sequential recommendation aims to predict the next item \(h_{n + 1}^u \in \mathcal{I}\) given a user \(u \in \mathcal{U}\) and item sequence \(H_u = (h_1^u, \ldots, h_n^u)\).
\negskip
\subsection{BSARec}
\begin{figure}[t]
    \centering
    \begin{minipage}[t]{0.55\linewidth}
        \centering
        \includegraphics[width=\linewidth]{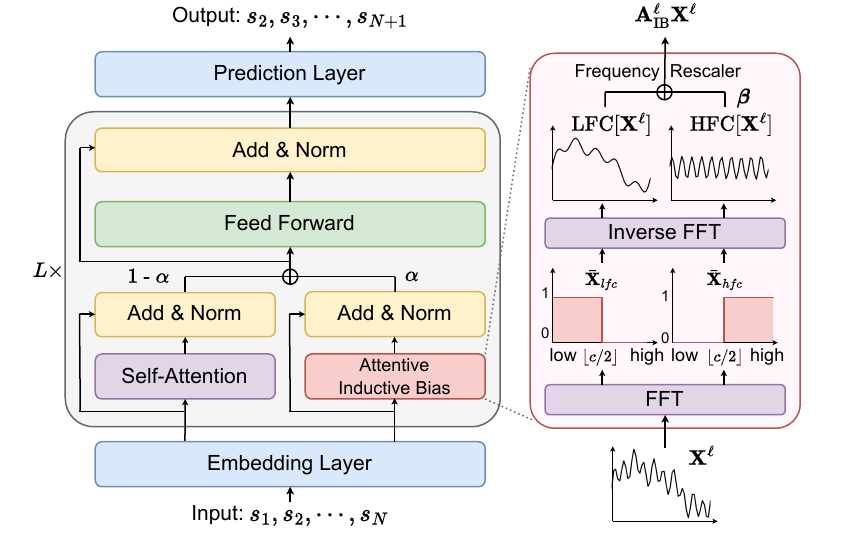}
        \caption{The BSARec architecture, reproduced from~\cite{DBLP:conf/aaai/Shin0WP24}.}
        \label{fig:arch}
    \end{minipage}
    \hfill
    \begin{minipage}[t]{0.43\linewidth}
        \centering
        \includegraphics[width=\linewidth]{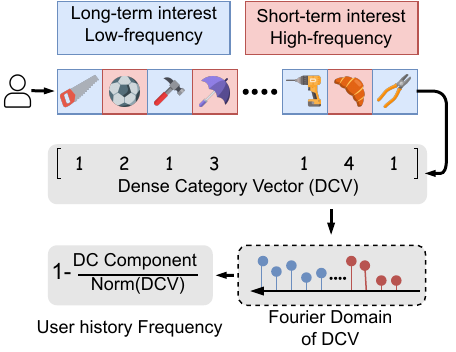}
        \caption{The calculation of the user history frequency, using the DC component.}
        \label{fig:pipeline}
    \end{minipage}
\end{figure}
 \citet{DBLP:conf/aaai/Shin0WP24} proposed the \ac{BSA} layer (see \cref{fig:arch}) to address the limited ability of attention to process short-term interests. Specifically, the attentive inductive bias module is designed as a supplement to the standard Transformer encoder~\cite{DBLP:conf/nips/VaswaniSPUJGKP17}.
The frequency rescaler decomposes an input \(x\) into a low-frequency component \(\operatorname{LFC_c(x)}\) and a high-frequency component \(\operatorname{HFC_c(x)}\), and rescales the high frequencies:
\begin{align}
    \operatorname{Rescale}_c(x) \coloneqq \operatorname{LFC_c}(x) + \beta \odot \operatorname{HFC_c}(x),
\end{align}
where \(\beta \in [0, \infty)^d\) is a learnable parameter and \(\odot\) is the Hadamard product (broadcast suitably).
\ac{BSA} combines the normalized output of the frequency rescaler with the output of the attention layer using a weighted sum \((1-\alpha)\operatorname{Attn}(X) + \alpha \operatorname{Rescale}_c(X)\), where \(X\) is a data matrix, \(\operatorname{Attn}\) is the self-attention operator, \(\alpha \in [0, 1]\) is a hyperparameter and all rescaling is applied per datapoint.
\subsection{Frequency Analysis of User Histories}
\label{section:scaled_dc_component}
By examining the distribution of user history frequencies in each dataset, we can analyze how effectively the models predict items from user histories that reflect short- and long-term interests. Intuitively, the signal of user history containing more long-term interests interactions will be `flatter'.
To further analyze models based on user interests, we propose the scaled DC component for frequency analysis (see \cref{fig:pipeline}) to measure how similar a user history \(H_u\) is to a flat distribution. Based on the mapping from items \(\mathcal{I}\) to a set of categories \(\mathcal{C}\), the user interaction history \(H_u\) can be converted to a category sequence \(\mathcal{C}_u\). Let \(m(u)\) be the number of unique categories occurring in \(\mathcal{C}_u\). Traverse the category sequence from left to right, each unique category in \(\mathcal{C}_u\) can be mapped to a number \(\ell \in \{1, 2, \ldots, m(u)\}\), resulting in a dense category vector \(d_u\). For example, if the category sequence \(\mathcal{C}_u = (\text{food}, \text{books}, \text{gadget}, \text{books})\), then \(d_u = (1, 2, 3, 2)\) in the new encoding. 
In the \ac{DFT}, the non-scaled \ac{DC} component \(\mathcal{F}(d_u)_1\) can be interpreted as the average of the signal.
Normalizing this component yields a connection to the cosine similarity: \(\mathcal{F}(d_u)_1/||d_u||_2 = \operatorname{similarity}(\mathbf{1}, d_u)\), where \(\mathbf{1}\) is a vector of ones.
This motivates us to define the \textbf{scaled DC component} as:
\begin{align}
    \operatorname{scaled DC}(x) \coloneqq 1 - \frac{\mathcal{F}(d_u)_1}{||d_u||_2} = 1 - \operatorname{similarity}(\mathbf{1}, d_u).
\end{align}
A user interaction history with a higher scaled DC component value can be interpreted as containing more high-frequency behavior.

\section{Reproducibility setup}
\header{Datasets} Experiments are conducted on Amazon datasets (Beauty, Sports and Outdoors, Toys and Games)~\cite{DBLP:journals/corr/abs-2403-03952}, ML-1M~\cite{DBLP:journals/tiis/HarperK16}, Yelp \cite{DBLP:journals/corr/Asghar16} and LastFM~\cite{DBLP:conf/ismir/Bertin-MahieuxEWL11} following  \citet{DBLP:conf/aaai/Shin0WP24}. Additionally, we evaluate model performance on the news dataset MIND~\cite{DBLP:conf/acl/WuQCWQLLXGWZ20}.
Descriptive statistics of the datasets are included in \cref{tab:datasets}.
\begin{table}[t]
    \caption{Statistics about the utilized datasets.}
    \centering
    \begin{adjustbox}{max width=0.9\textwidth}
    \begin{tabular}{lrrrrr}
    \toprule
         \textbf{Dataset} & \textbf{Users} & \textbf{Items} & \textbf{Interactions} & \textbf{Avg. Length} & \textbf{Sparsity} \\
         \midrule
         Beauty~\cite{DBLP:journals/corr/abs-2403-03952} & 22,363 & 12,101 & 198,502 & 8.9 & 99.93\%\\
         ML-1M~\cite{DBLP:journals/tiis/HarperK16} & 6,041 & 3,417 & 999,611 & 165.5 & 95.16\%\\
         Sports and Outdoors~\cite{DBLP:journals/corr/abs-2403-03952} & 25,598 & 18,357 & 296,337 & 8.3 & 99.95\%\\
         Toys and Games~\cite{DBLP:journals/corr/abs-2403-03952} & 19,412 & 11,924 & 167,597 & 8.6 & 99.93\%\\
         LastFM~\cite{DBLP:conf/ismir/Bertin-MahieuxEWL11} & 1,090 & 3,646 & 52,551 & 48.2 & 98.68\% \\
         Yelp~\cite{DBLP:journals/corr/Asghar16} & 30,431 & 20,033 & 316,354 & 10.4 & 99.95\%\\
        MIND~\cite{DBLP:conf/acl/WuQCWQLLXGWZ20} & 91,935 & 48,204 & 230,117 & 21.0 & 99.99\%\\
      \bottomrule  
    \end{tabular}
    \end{adjustbox}
    \label{tab:datasets}
\end{table}

\header{Models}
We compare BSARec to five \ac{SR} models from two model classes, following \citet{DBLP:conf/aaai/Shin0WP24}.
\begin{enumerate*}[label=(\roman*)]
\item \textbf{The models using \ac{DSP}.} FEARec~\cite{DBLP:conf/sigir/DuYZQZ0LS23} uses a hybrid attention layer with one attention layer in the Fourier domain and one attention layer in the time domain. FMLPRec~\cite{DBLP:conf/www/ZhouYZW22} stacks feedforward networks with learnable filters in the Fourier domain.
\item \textbf{The models \textit{without} DSP components}. SASRec~\cite{DBLP:conf/icdm/KangM18} stacks blocks consisting of one self-attention layer and one point-wise MLP. BERT4Rec~\cite{DBLP:conf/cikm/SunLWPLOJ19} uses bidirectional self-attention and employs an item masking training task. DuoRec~\cite{DBLP:conf/wsdm/QiuHYW22} extends SASRec with contrastive learning to uniformize item embeddings distribution.

\end{enumerate*}

\header{Metrics} Following~\cite{DBLP:conf/wsdm/QiuHYW22, DBLP:conf/sigir/DuYZQZ0LS23, DBLP:journals/ipm/SuCL25, DBLP:conf/aaai/Shin0WP24, DBLP:conf/www/ZhouYZW22}, we use \ac{NDCG}~\cite{DBLP:conf/colt/WangWLHL13} and \ac{HR}~\cite{DBLP:conf/valuetools/FofackNNT12} to evaluate model performance. Each utilizes a top-\(k\) cutoff for the recommendations, denoted with the postfix `@k'. 

\header{Implementation Details} Following~\cite{DBLP:conf/aaai/Shin0WP24}, the dataset split is determined by the positions in the user history, with the final three indices of each subsequence used as labels for training, validation, and test, respectively. For each subsequence, the encoded sequence length  \(\ell =50\) most recent items in the user's history, excluding the target item, are used for training and inference.  Histories containing fewer items than \(\ell\) are zero padded. The hyperparameters reported by~\cite{DBLP:conf/aaai/Shin0WP24} were re-used, we tuned models for novel dataset MIND using a grid search over the same grid as~\cite{DBLP:conf/aaai/Shin0WP24}: $\alpha \in \{0.1, 0.3, 0.5, 0.7, 0.9\}$, $c \in \{1,3,5,7,9\}$, learning rate $ \eta \in \{5 \cdot 10^{-4}, 1\cdot 10^{-3}\}$, and the number of attention heads $h\in\{1,2,4\}$.
The hyperparameters reported by  \cite{DBLP:conf/aaai/Shin0WP24} and the hyperparameters from our search are re-used throughout all experiments due to resource constraints.
Experiments with the \ac{DWT} are performed using the Haar wavelet, as a hyperparameter search over various wavelet types, including variants of Biorthogonal, Coiflet, Meyer wavelet, Symlet and Daubechies wavelets. 
However, the choice of wavelet was not found to significantly impact performance. 
All experiments were carried out with five random seeds on an A100 GPU, using the implementation from the official repository of~\cite{DBLP:conf/aaai/Shin0WP24}; average results are reported. 
We use a \(t\)-test to determine statistically significant differences \((p < 0.05)\) between BSARec and the best performing baseline.
Our repository is available on Github via \githuburl.

\section{Experiments and results}
\subsection{\cref*{rq:reproduction}: Reproduction}\label{section:reproduction}
We first reproduce BSARec and five \ac{SR} models to address \cref{rq:reproduction}. Results on seven \ac{SR} benchmarks are shown in~\cref{tab:main}. It can be observed that
\begin{enumerate*}[label=(\roman*)]
\item BSARec achieves the best scores on the datasets LastFM, ML-1M and `Toy and Games', with a large improvement on LastFM. This improvement may be due to the dataset containing users with abrupt interest changes, i.e., shifts in music genres. Furthermore, BSARec demonstrates greater improvements over the baselines on less sparse datasets, as seen in \cref{tab:datasets}.

\item In contrast to the reported results by  \cite{DBLP:conf/aaai/Shin0WP24}, other baselines significantly outperform BSARec on the Yelp dataset across all metrics, and on the `Sports and Outdoors' dataset for NDCG. As noted by \citet{DBLP:conf/aaai/Shin0WP24}, BSARec does not incorporate contrastive learning, which may account for its lower performance relative to FEARec and DuoRec in~\cref{tab:main}. Discrepancies from the original results may further arise from differences in hardware configurations or the unspecified number of random seeds.

\item To generalize the results of BSARec and validate the findings, an additional dataset MIND was employed. 
BSARec slightly underperforms other baselines on the MIND dataset. However, these differences are not significant. 
\end{enumerate*}

\begin{figure}[t]
    \centering
    \includegraphics[width=0.9\linewidth]{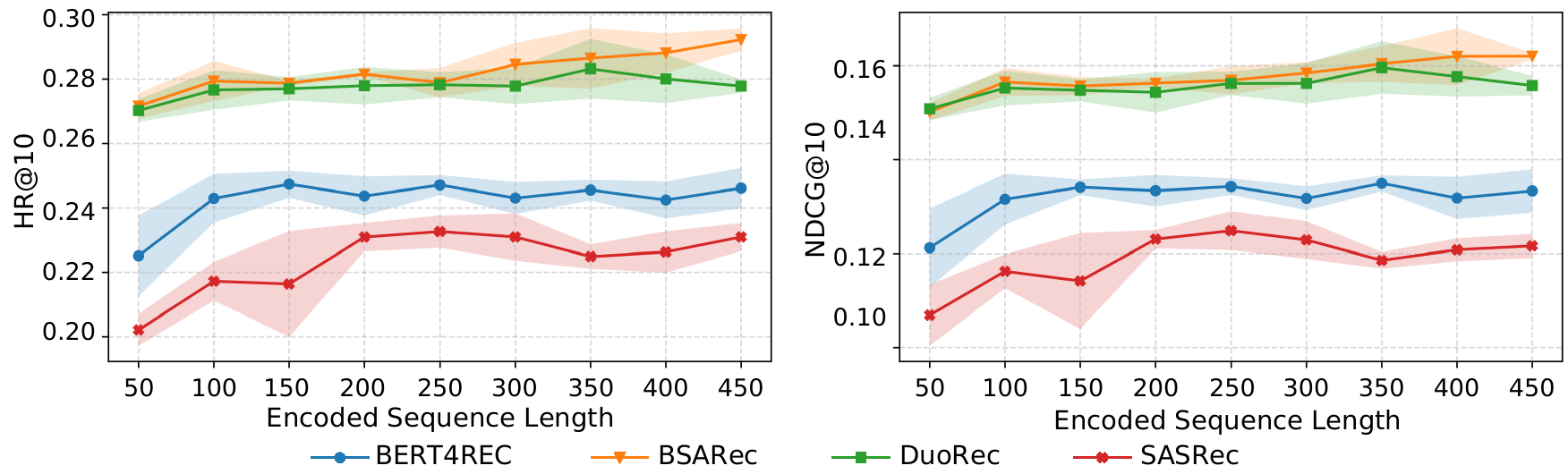}
    \caption{Performance comparison on the ML-1M dataset for various sequence lengths. The error bars denote standard deviations.}
    \label{fig:small_length_comparison}
\end{figure}

\header{Effect of Longer Inputs} To examine the performance of BSARec and other baselines on longer encoded sequence lengths \(\ell\), we extend \(\ell\) from 50 to 450 in steps of 50. 
We conduct experiments on ML-1M, which contains user histories of 165.5 items on average. Results are shown in~\cref{fig:small_length_comparison}. We summarize our observations below: 
\begin{enumerate*}[label=(\roman*)]
\item 
Across different encoded sequence lengths, the performance of all methods consistently maintains the ranking observed in~\cref{tab:main}.
\item All models perform better when increasing \(\ell\) from 50 to 100, likely due to the availability of more user history items in the encoded sequence.
\item The performance of BSARec slightly increases for encoded sequence lengths \(\ell \geq 300\), though with larger standard deviations.
\end{enumerate*}
These observations suggest that the \ac{DFT} is effective in handling longer contexts. 

\begin{table*}[t!]
    \centering
    \caption{Results on seven \ac{SR} benchmarks. The best and second-best scores are shown in \textbf{bold} and \underline{underlined}, respectively. 
    Statistically significant differences between BSARec and the best baseline (\texttt{Diff}) are marked with \textsuperscript{*}.
    }
    \begin{adjustbox}{max width=\textwidth}
    \label{tab:main}
    \renewcommand{\arraystretch}{0.9}
\begin{tabular}{llccccccc}
\toprule
\textbf{Dataset} & \textbf{Metric} &\textbf{SASRec} & \textbf{BERT4Rec} & \textbf{FMLPRec} & \textbf{DuoRec} & \textbf{FEARec} & \textbf{BSARec} & \textbf{Diff.} \\
\midrule
\multirow{6}{*}{\centering Beauty} & HR@5 & 0.0297 & 0.0341 & 0.0347 & \underline{0.0517} & \textbf{0.0518} & 0.0508 & -1.87\textsuperscript{*} \\
 & HR@10 & 0.0466 & 0.0537 & 0.0535 & 0.0705 & \textbf{0.0712} & \underline{0.0712} & -0.13~ \\
 & HR@20 & 0.0703 & 0.0802 & 0.0793 & 0.0961 & \underline{0.0971} & \textbf{0.0987} & 1.62~ \\
 & NDCG@5 & 0.0191 & 0.0221 & 0.0227 & \underline{0.0372} & \textbf{0.0372} & 0.0368 & -1.23~ \\
 & NDCG@10 & 0.0246 & 0.0284 & 0.0288 & 0.0432 & \textbf{0.0435} & \underline{0.0433} & -0.40~ \\
 & NDCG@20 & 0.0305 & 0.0351 & 0.0353 & 0.0497 & \underline{0.0500} & \textbf{0.0503} & 0.49~ \\
\midrule
\multirow{6}{*}{\centering LastFM} & HR@5 & 0.0362 & 0.0294 & 0.0393 & \underline{0.0407} & 0.0400 & \textbf{0.0477} & 17.12\textsuperscript{*} \\
 & HR@10 & 0.0546 & 0.0495 & \underline{0.0591} & 0.0547 & 0.0543 & \textbf{0.0703} & 18.94\textsuperscript{*} \\
 & HR@20 & 0.0837 & 0.0754 & \underline{0.0839} & 0.0829 & 0.0785 & \textbf{0.1015} & 21.01\textsuperscript{*} \\
 & NDCG@5 & 0.0244 & 0.0191 & 0.0253 & \underline{0.0294} & 0.0285 & \textbf{0.0332} & 12.99~ \\
 & NDCG@10 & 0.0303 & 0.0256 & 0.0316 & \underline{0.0338} & 0.0331 & \textbf{0.0405} & 19.73\textsuperscript{*} \\
 & NDCG@20 & 0.0376 & 0.0320 & 0.0379 & \underline{0.0409} & 0.0392 & \textbf{0.0483} & 18.07\textsuperscript{*} \\
\midrule
\multirow{6}{*}{\centering ML-1M} & HR@5 & 0.1263 & 0.1474 & 0.1302 & \underline{0.1851} & 0.1814 & \textbf{0.1888} & 2.02~ \\
 & HR@10 & 0.2008 & 0.2234 & 0.2090 & 0.2678 & \underline{0.2700} & \textbf{0.2746} & 1.73~ \\
 & HR@20 & 0.3052 & 0.3263 & 0.3187 & 0.3731 & \underline{0.3737} & \textbf{0.3814} & 2.05\textsuperscript{*} \\
 & NDCG@5 & 0.0815 & 0.0974 & 0.0832 & \underline{0.1245} & 0.1218 & \textbf{0.1263} & 1.45~ \\
 & NDCG@10 & 0.1054 & 0.1220 & 0.1085 & \underline{0.1512} & 0.1503 & \textbf{0.1539} & 1.80\textsuperscript{*} \\
 & NDCG@20 & 0.1316 & 0.1479 & 0.1361 & \underline{0.1777} & 0.1764 & \textbf{0.1808} & 1.74\textsuperscript{*} \\
\midrule
\multirow{6}{*}{\centering Sports and Outdoors} & HR@5 & 0.0137 & 0.0164 & 0.0163 & \textbf{0.0243} & \underline{0.0238} & 0.0237 & -2.40~ \\
 & HR@10 & 0.0221 & 0.0275 & 0.0256 & \textbf{0.0351} & 0.0346 & \underline{0.0349} & -0.54~ \\
 & HR@20 & 0.0350 & 0.0432 & 0.0397 & \underline{0.0502} & 0.0500 & \textbf{0.0504} & 0.35~ \\
 & NDCG@5 & 0.0087 & 0.0106 & 0.0106 & \textbf{0.0171} & \underline{0.0168} & 0.0164 & -4.07\textsuperscript{*} \\
 & NDCG@10 & 0.0114 & 0.0141 & 0.0136 & \textbf{0.0206} & \underline{0.0203} & 0.0200 & -2.77\textsuperscript{*} \\
 & NDCG@20 & 0.0146 & 0.0181 & 0.0171 & \textbf{0.0244} & \underline{0.0242} & 0.0239 & -1.95\textsuperscript{*} \\
\midrule
\multirow{6}{*}{\centering Toys and Games} & HR@5 & 0.0340 & 0.0283 & 0.0378 & \underline{0.0481} & 0.0473 & \textbf{0.0502} & 4.35\textsuperscript{*} \\
 & HR@10 & 0.0476 & 0.0429 & 0.0531 & \underline{0.0650} & 0.0642 & \textbf{0.0672} & 3.38\textsuperscript{*} \\
 & HR@20 & 0.0654 & 0.0654 & 0.0729 & \underline{0.0881} & 0.0878 & \textbf{0.0917} & 4.03\textsuperscript{*} \\
 & NDCG@5 & 0.0231 & 0.0193 & 0.0257 & \underline{0.0358} & 0.0352 & \textbf{0.0369} & 3.10\textsuperscript{*} \\
 & NDCG@10 & 0.0275 & 0.0240 & 0.0307 & \underline{0.0412} & 0.0406 & \textbf{0.0424} & 2.77\textsuperscript{*} \\
 & NDCG@20 & 0.0320 & 0.0296 & 0.0357 & \underline{0.0471} & 0.0466 & \textbf{0.0485} & 3.15\textsuperscript{*} \\
\midrule
\multirow{6}{*}{\centering Yelp} & HR@5 & 0.0122 & 0.0190 & 0.0152 & \textbf{0.0203} & \underline{0.0192} & 0.0182 & -10.38\textsuperscript{*} \\
 & HR@10 & 0.0212 & \underline{0.0334} & 0.0259 & \textbf{0.0340} & 0.0330 & 0.0309 & -9.31\textsuperscript{*} \\
 & HR@20 & 0.0357 & 0.0559 & 0.0438 & \textbf{0.0569} & \underline{0.0561} & 0.0518 & -8.96\textsuperscript{*} \\
 & NDCG@5 & 0.0076 & \underline{0.0118} & 0.0095 & \textbf{0.0127} & 0.0118 & 0.0114 & -10.85\textsuperscript{*} \\
 & NDCG@10 & 0.0105 & \underline{0.0164} & 0.0129 & \textbf{0.0172} & 0.0162 & 0.0154 & -10.08\textsuperscript{*} \\
 & NDCG@20 & 0.0142 & \underline{0.0221} & 0.0174 & \textbf{0.0229} & 0.0220 & 0.0207 & -9.65\textsuperscript{*} \\
 \midrule
\multirow{6}{*}{\centering MIND} & HR@5 & 0.0690 & \textbf{0.1143} & 0.0624 & {0.1121} & \underline{0.1141} & 0.1087 & -4.87 \\
 & HR@10 & 0.1068 & \textbf{0.1739} & 0.0987 & {0.1692} & \underline{0.1710} & 0.1633 & -6.11~ \\
 & HR@20 & 0.1586 & \textbf{0.2506} & 0.1529 & {0.2429} & \underline{0.2456} & 0.2354 & -6.08~ \\
 & NDCG@5 & 0.0462 & \underline{0.0755} & 0.0412 & {0.0743} & \textbf{0.0760} & 0.0724 & -4.74~ \\
 & NDCG@10 & 0.0584 & \textbf{0.0947} & 0.0529 & {0.0926} & \underline{0.0943} & 0.0899 & -5.07~ \\
 & NDCG@20 & 0.0714 & \textbf{0.1141} & 0.0665 & {0.1112} & \underline{0.1131} & 0.1081 & -5.24~ \\ 
\bottomrule
\end{tabular}
\end{adjustbox}
\end{table*}

\subsection{\cref*{rq:history_frequency}: Analysis of Low- and High-frequency Behavior}\label{sec:lowhighres}
To answer \cref{rq:history_frequency}, we compare performance of \ac{SR} methods when considering 
\begin{enumerate*}[label=(\arabic*)]
\item user history frequencies\footnote{In our paper, frequency refers to the frequency of a signal (the user’s history) in the Fourier domain. We use the term occurrence count or the number of occurrences to denote how many times a specific category appears in the user’s past interactions.} and
\item the number of occurrences of the target category in the user history, 
\end{enumerate*}
as two indicators of long- or short-term interest.

\begin{figure}[t]
    \centering
    \includegraphics[width=0.78\linewidth]{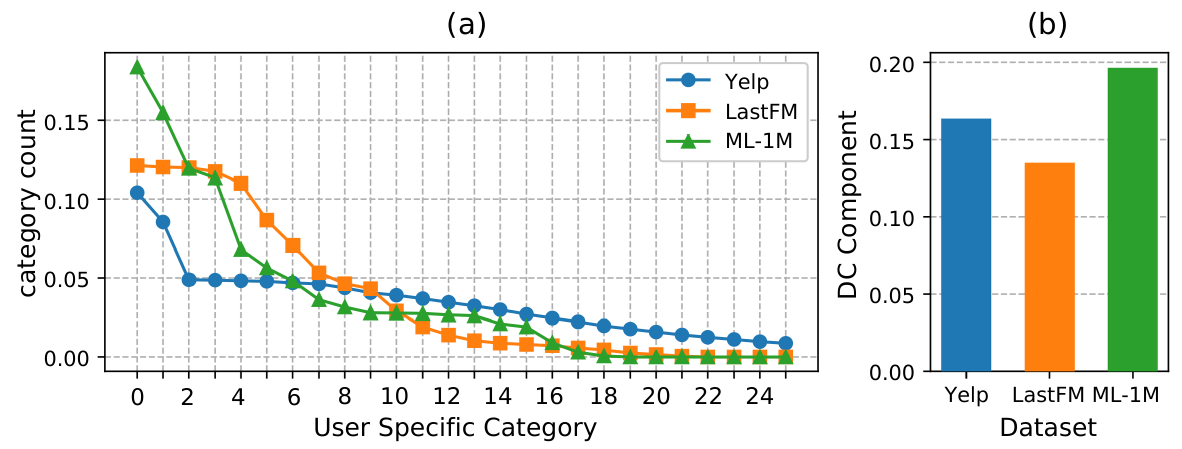}
    \caption{(a) Statistics of item category occurrences in users' history, x-axis is capped at 25 due to the long-tailed distribution of category counts. (b) The average scaled DC component of user histories per dataset. 
    }
    \label{fig:dataset_frequency}
\end{figure}

\header{User Frequencies} First, we analyze the statistical distribution of the occurrence counts of item categories in users’ purchase histories in the dataset, and compute the average scaled DC component for all users as described in~\cref{section:scaled_dc_component}. We select the Yelp, LastFM, and ML-1M datasets in our experiment since they are anticipated to exhibit substantial variation in user behavior over time, making them well-suited for analyzing both low-frequency and high-fre\-quency interaction patterns.
In \cref{fig:dataset_frequency}(a), we rank categories in each user's history based on their occurrence proportion, and compute the average category occurrence proportion at each position (x-axis: from 0 to 25) across all the users. 
\cref{fig:dataset_frequency}(b) shows the scaled \ac{DC} component of three datasets, which indicates how much short- and long-term interests are included in each dataset. 
We observe that ML-1M has the largest amount of short-term interests, followed by Yelp and LastFM. This is also evident in \cref{fig:dataset_frequency}(a), where each user’s history in ML-1M typically contains few categories, while others appear more sporadically, suggesting that user histories are dominated by high-frequency signals. 
LastFM and Yelp exhibit a sharper distribution of category occurrences per user history, indicating that these histories contain stronger low-frequency signals.

\begin{figure}[!t]
    \centering
    \includegraphics[width=0.9\linewidth]{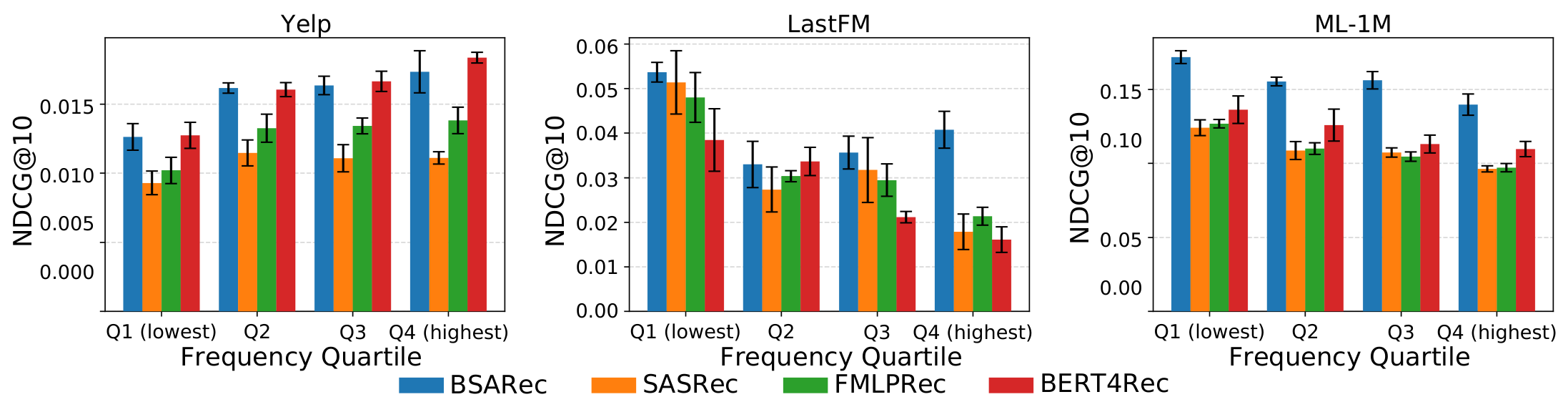}
    \caption{Performance of recommendation methods across user groups with different DC history frequencies on Yelp, LastFM, and ML-1M. 
}
    \label{fig:high_low_performance}
\end{figure}

Furthermore, we compare the performance of BSARec and three baseline models across user groups on Yelp, LastFM and ML-1M datasets. Users are grouped by the quartile of \ac{DC} frequency of their histories: Q1 represents those with the lowest values, whose histories are dominated by long-term interests, while Q4 corresponds to the highest values, reflecting that users have more short-term interest shifts.
From \cref{fig:high_low_performance}, we observe that:
\begin{enumerate*}[label=(\roman*)]
\item BSARec outperforms all baseline models across all user groups on the LastFM and ML-1M datasets, and achieves comparable results to BERT4Rec on Yelp.

\item The performance of these four \ac{SR} models shows an increasing trend on Yelp dataset when the users' history frequency increase. However, the trend is opposite on ML-1M dataset.

\item Notably, on the LastFM dataset, BSARec’s performance improves from $Q_2$ to $Q_4$, while other models show a declining trend. This could indicate improved recommendation performance for high-frequency users on LastFM for BSARec.
\end{enumerate*}

\begin{figure}[t]
    \centering
    \includegraphics[width=0.9\linewidth]{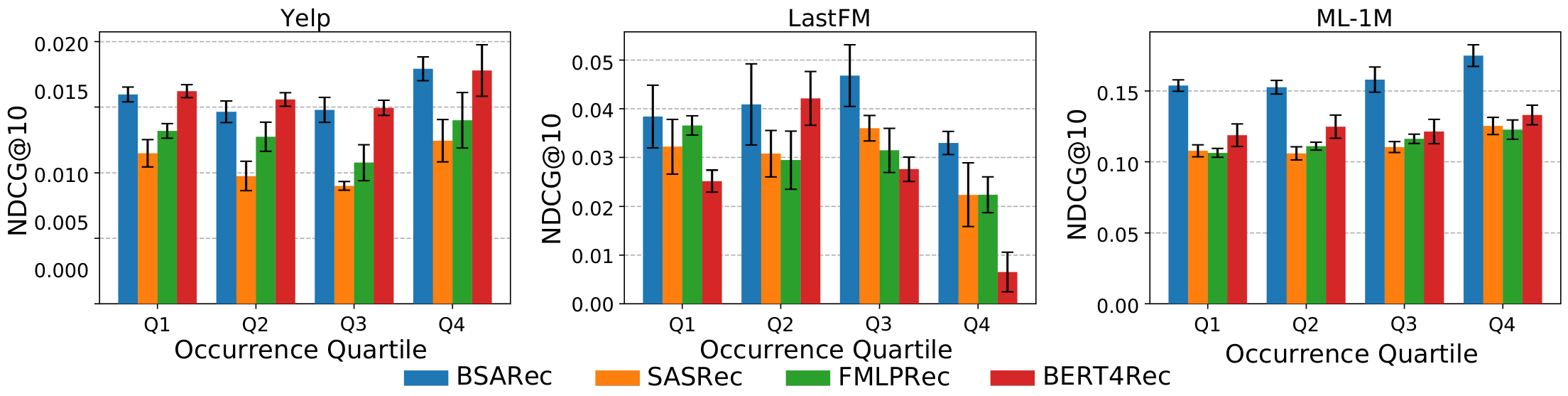}
    \caption{Performance of recommendation methods on Yelp, LastFM, and ML-1M, grouped by the occurrence of the target item's category in the user history. 
    }
    \label{fig:new_freq}
\end{figure}

\negskip
\header{Category Occurrence of Target Item}
We also split the users based on the category occurrence of the target item. Specifically, the item categories in each user's history are divided into quartiles from low to high occurrence. $Q_1$ represents the user group whose target item belongs to the most common categories in their interaction histories, whereas $Q_4$ corresponds to the group whose target items belong to the rarest categories. 
\Cref{fig:new_freq} shows BSARec and baseline results when grouped by target items. We observe that:
\begin{enumerate*}[label=(\roman*)]
\item BSARec achieves the best performance across all groups on the LastFM and ML-1M datasets, and performs comparably to BERT4Rec on the Yelp dataset.
\item On Yelp and ML-1M datasets, all the models achieve the best performance when predicting the target items belonging to the rarest categories ($Q_4$). In contrast, on the LastFM dataset, all models yield the lowest performance for $Q_4$. 
\end{enumerate*}

Through a two-perspective analysis of user history and target items, we observe that \ac{SR} methods exhibit distinct performance patterns for short-term and long-term interest groups across different datasets. Notably, on the LastFM and ML-1M datasets, BSARec achieves a significant performance advantage over other \ac{SR} methods on the high-frequency data group.

\subsection{\cref*{rq:dsp}: Signal Processing Method}\label{RQ4:FourierVSWavelet}
\begin{figure}[t]
    \centering
    \includegraphics[width=0.9\linewidth]{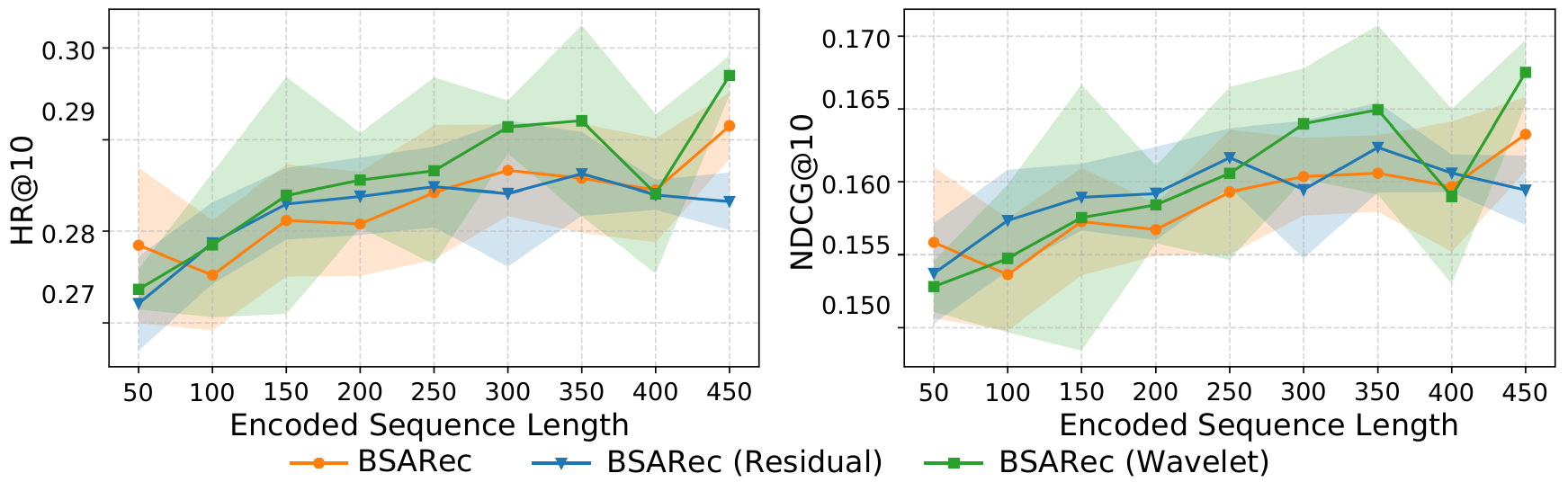}
    \caption{Performance comparison on the ML-1M dataset for BSARec (Fourier), BSARec (wavelet), and BSARec (Residual connection).}
    \label{fig:small_length_comparison_freq}
\end{figure}

\negskip
To answer \cref{rq:dsp}, we explore an alternative \ac{DSP} method by replacing the \ac{DFT} in BSARec with the \ac{DWT} to examine the impact of the chosen \ac{DSP} method. Results comparing the \ac{DFT} and \ac{DWT} in BSARec are visualized in \cref{tab:signal-domain}. 
Furthermore, we extend to different encoded sequence lengths $\ell$ and include an ablation study using a residual connection.
\cref{fig:small_length_comparison_freq} compares BSARec architecture with the \ac{DFT}, the \ac{DWT}, and a residual connections on ML-1M dataset.

\negskip
\header{Choice of DSP Method}
\begin{enumerate*}[label=(\roman*)]
\item In~\cref{tab:signal-domain}, Fourier transform yields higher recommendation scores across all datasets, but few differences are significant.

\item From~\cref{fig:small_length_comparison_freq}, both DSP methods demonstrate an increasing trend in recommendation performance. In ML-1M, the average user history length is 165.5, hence larger $\ell$ likely yields higher scores by encoding more history.

\item There is no clear advantage for either method; although performance is generally higher for the \ac{DWT} on longer sequences, the results consistently remain within one standard deviation of each other.

\end{enumerate*}

\negskip
\begin{table*}[t]
\centering
\caption{Performance comparison of BSARec with the Fourier transform (F) and the wavelet transform (W). The best average score is \textbf{bolded}. Statistically significant differences between F and W (\texttt{Diff}) are marked with \textsuperscript{*}}
\label{tab:signal-domain}
\begin{adjustbox}{max width=\textwidth}
\begin{tabular}{lccccccccc}
\toprule
& \multicolumn{3}{c}{\textbf{Beauty}} 
& \multicolumn{3}{c}{\textbf{LastFM}} 
& \multicolumn{3}{c}{\textbf{ML-1M}} \\
\cmidrule(lr){2-4} \cmidrule(lr){5-7} \cmidrule(lr){8-10}
\textbf{Metric} & F & W & Diff (\%) & F & W & Diff (\%) & F & W & Diff (\%) \\
\midrule
HR@5     & \textbf{0.0508} & 0.0508 & 0.05~   & \textbf{0.0477} & 0.0440 & 8.33~ & \textbf{0.1888} & 0.1852 & {1.95}\textsuperscript{*} \\
HR@10    & \textbf{0.0712} & 0.0705 & 0.86~   & \textbf{0.0703} & 0.0642 & 9.43\textsuperscript{*} & \textbf{0.2746} & 0.2736 & 0.36~ \\
HR@20    & \textbf{0.0987} & 0.0978 & 0.92~   & \textbf{0.1015} & 0.0989 & 2.60~ & \textbf{0.3814} & 0.3773 & 1.10~ \\
NDCG@5   & \textbf{0.0368} & 0.0363 & 1.45~   & \textbf{0.0332} & 0.0311 & 6.77~ & \textbf{0.1263} & 0.1244 & 1.55~ \\
NDCG@10  & \textbf{0.0433} & 0.0426 & {1.62}\textsuperscript{*} 
         & \textbf{0.0405} & 0.0376 & 7.79~ & \textbf{0.1539} & 0.1528 & 0.72~ \\
NDCG@20  & \textbf{0.0503} & 0.0495 & 1.58~   & \textbf{0.0483} & 0.0462 & 4.63~ & \textbf{0.1808} & 0.1790 & 1.01~ \\
\bottomrule
\end{tabular}
\end{adjustbox}
\end{table*}

\header{Residual Connection} 
To verify whether performance improvements can be attributed to the use of DSP methods, we remove the Fourier transform in the frequency rescaler, effectively transforming the rescaler into a residual connection~\cite{DBLP:conf/cvpr/HeZRS16}, and compare results in \cref{fig:small_length_comparison_freq}. 
We observe that:  
\begin{enumerate*}[label=(\roman*)]
\item BSARec with residual connection marginally outperforms BSARec on sequences shorter than 250 items, though the scores generally remain within one standard deviation of each other.

\item The performances of BSARec with and without the DSP method are comparable, given the substantial overlap of their error bands.
However, since the hyperparameters for \(\ell = 50\) were re-used, BSARec may not show optimal performance on longer encoded sequences.
\end{enumerate*}

\subsection{\cref*{rq:padding}: The Influence of Padding}
To answer \cref{rq:padding}, we analyze the frequency decomposition of the user history representation and study different padding methods.
The used padding methods are: \begin{enumerate*}[label=(\arabic*)]
    \item zero,
    \item cyclic,
    \item reflect and
    \item symmetric.
\end{enumerate*}
Padding is added to the left of the actual user history.
Zero padding uses a zero token, while cyclic padding repeats the input sequence.
Reflect padding and symmetric padding both mirror the sequence at the beginning of the user history to generate padding, where reflect padding omits the first item of the history from the mirrored sequence.

\header{Qualitative Signal Representation} An input representation \(x\) can be decomposed into low- and high-frequency components \(\operatorname{LFC}_c(x)\) and \(\operatorname{HFC}_c (x)\), respectively.
\cref{fig:user9_lastfm} shows the \(L_2\) norm of both components for each item in an encoded sequence. We make the following observation:
\begin{enumerate*}[label=(\roman*)]
\item When using zero padding, the low-frequency component of the input is dominated by the padding embeddings, reflect padding does not exhibit this effect. This observation can be explained by zero padding acting as low-frequency signal.
Consequently, high-frequency components consist mostly of the non-padding input signal, limiting the ability of the frequency rescaler in BSARec to identify high-frequency user interests.
\item Reflect padding results in a different LFC and HFC decomposition, as the non-constant padding prevents it from dominating the low-frequency component.
\end{enumerate*}

\begin{figure}[t]
    \includegraphics[width=0.9\linewidth]{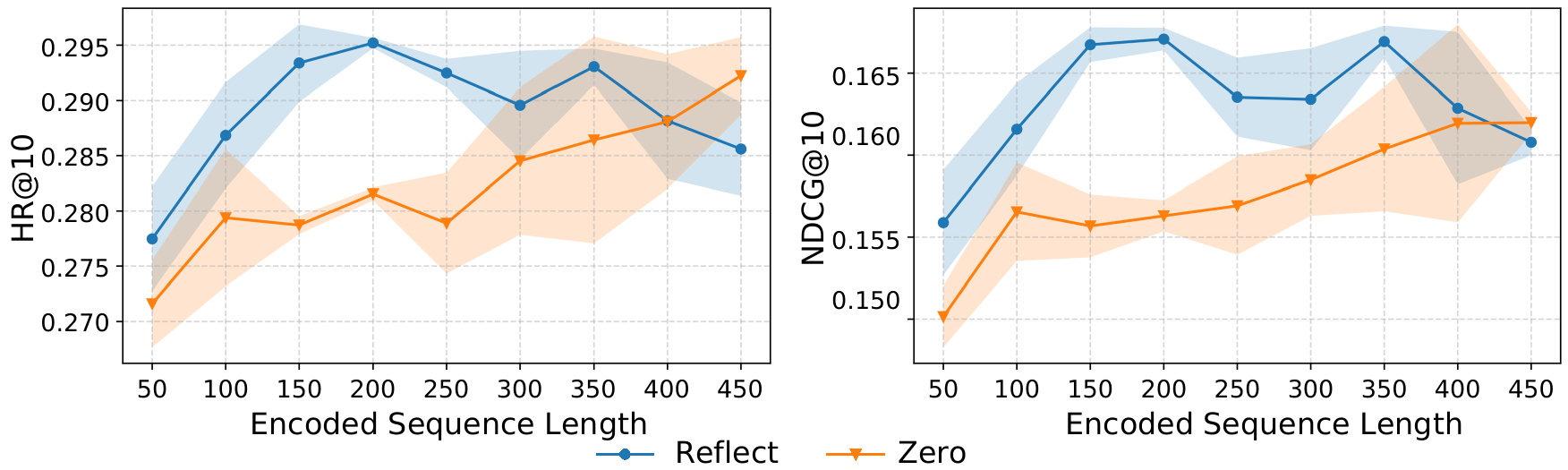}
    \centering
    \caption{Comparison between different selections of the maximum sequence length hyperparameters for ML-1M between BSARec with zero and reflect padding.}
    \label{fig:seq_len_comparison}
\end{figure}

\header{Quantitive Analysis} We compare padding methods in \cref{tab:padding}.
\begin{enumerate*}[label=(\roman*)]
\item When using reflect padding, the performance on the LastFM and Yelp datasets increases significantly compared to zero padding.
The difference in results for ML-1M is smaller, likely due to its high average sequence length (165.5). Since the encoded sequence length is fixed at 50, many user histories require no padding.
\item Generally, the cyclic and symmetric padding methods perform similar to the reflect method, showing improved performance over zero padding. 

\cref{fig:seq_len_comparison} compares reflect and zero padding for varying encoded sequence lengths on ML-1M.
\item we observe that reflect padding shows improved performance up to an encoded sequence length of \(\ell = 200\), after which the gains diminish for longer sequences. This may be due to the increased proportion of padding that is added for longer sequences, leading to diminishing returns~\citep{DBLP:conf/iclr/AlsallakhKMYR21}.

\end{enumerate*}

\begin{table*}[t]
    \centering
    \begin{minipage}[b]{0.49\linewidth}
        \centering
        \begin{adjustbox}{max width=0.90\textwidth}
    \begin{tabular}{lllccccc}
    \toprule
    \textbf{Dataset} & \textbf{Metric} &\textbf{Zero} & \textbf{Cyclic} & \textbf{Reflect} & \textbf{Symmetric} & \textbf{Diff.} \\
    \midrule
    \multirow{6}{*}{\centering LastFM} & HR@5 & 0.0497 & 0.0479 & \textbf{0.0499} & 0.0490 & -0.37~ \\
     & HR@10 & 0.0686 & 0.0730 & \textbf{0.0760} & 0.0701 & -9.66\textsuperscript{*} \\
     & HR@20 & 0.1017 & 0.1073 & \textbf{0.1114} & 0.1072 & -8.73\textsuperscript{*} \\
     & NDCG@5 & \textbf{0.0347} & 0.0329 & 0.0345 & 0.0346 & 0.42~ \\
     & NDCG@10 & 0.0408 & 0.0411 & \textbf{0.0430} & 0.0414 & -5.15~ \\
     & NDCG@20 & 0.0490 & 0.0497 & \textbf{0.0518} & 0.0507 & -5.47\textsuperscript{*} \\
    \midrule
    \multirow{6}{*}{\centering ML-1M} & HR@5 & 0.1892 & 0.1946 & \textbf{0.1947} & 0.1945 & -2.82~ \\
     & HR@10 & 0.2784 & 0.2807 & \textbf{0.2817} & 0.2801 & -1.14~ \\
     & HR@20 & 0.3832 & 0.3864 & \textbf{0.3882} & 0.3860 & -1.27~ \\
     & NDCG@5 & 0.1271 & \textbf{0.1317} & 0.1315 & 0.1306 & -3.44~ \\
     & NDCG@10 & 0.1558 & 0.1594 & \textbf{0.1595} & 0.1582 & -2.28~ \\
     & NDCG@20 & 0.1822 & 0.1860 & \textbf{0.1863} & 0.1849 & -2.18~ \\
    \midrule
    \multirow{6}{*}{\centering Yelp} & HR@5 & 0.0178 & 0.0187 & \textbf{0.0193} & 0.0190 & -7.43\textsuperscript{*} \\
     & HR@10 & 0.0310 & 0.0323 & \textbf{0.0332} & 0.0328 & -6.58\textsuperscript{*} \\
     & HR@20 & 0.0513 & 0.0548 & \textbf{0.0556} & 0.0554 & -7.77\textsuperscript{*} \\
     & NDCG@5 & 0.0112 & 0.0117 & \textbf{0.0121} & 0.0119 & -7.79\textsuperscript{*} \\
     & NDCG@10 & 0.0154 & 0.0161 & \textbf{0.0166} & 0.0164 & -7.22\textsuperscript{*} \\
     & NDCG@20 & 0.0205 & 0.0217 & \textbf{0.0222} & 0.0220 & -7.77\textsuperscript{*} \\
    \bottomrule
    \end{tabular}
    \end{adjustbox}
    \captionsetup{skip=10pt}
    \caption{Performance of BSARec with different padding methods.
        Best scores are in \textbf{bold}.
        Statistically significant changes between Zero and the best Padding method (\texttt{Diff}) are marked with \textsuperscript{*} }
        \label{tab:padding}
    \end{minipage}
    \hfill
    \begin{minipage}[b]{0.49\linewidth}
    \centering
        \includegraphics[width=1\linewidth]{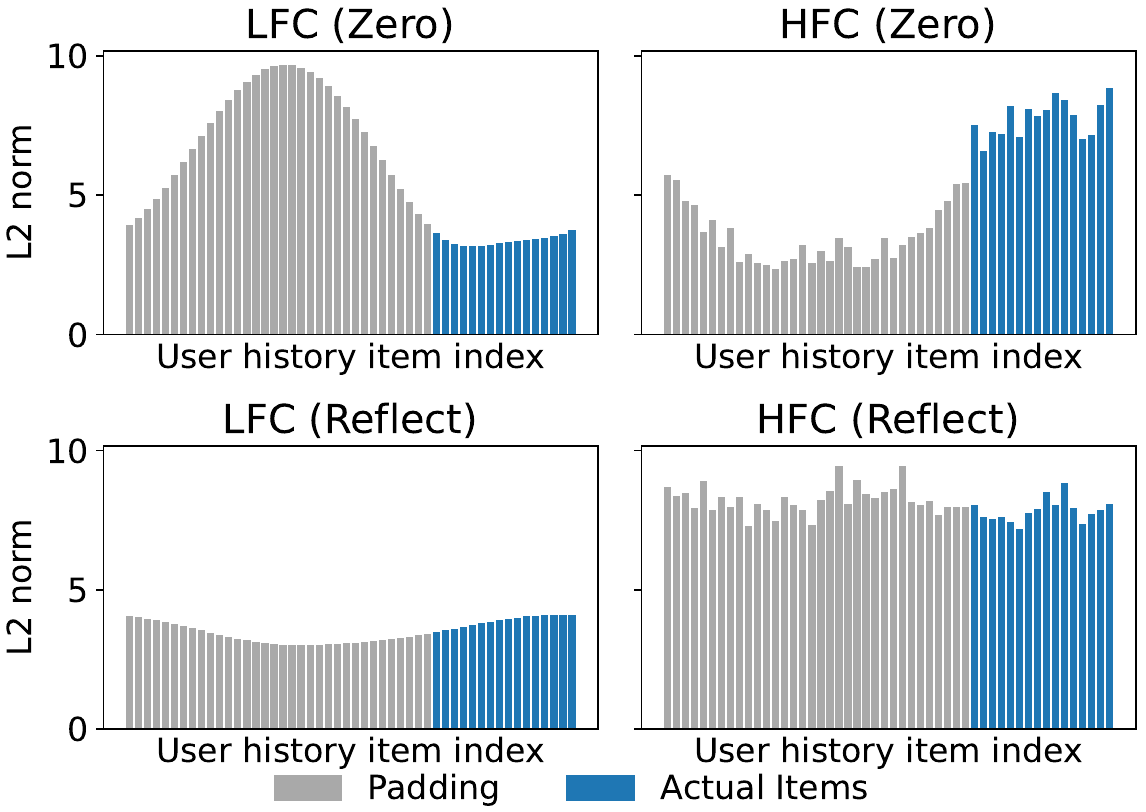}
    \captionof{figure}{\(L_2\) norm of low- and high-frequency components of the item embeddings appearing in a user’s history from the LastFM dataset, comparing BSARec with different padding.}
    \label{fig:user9_lastfm}
    \end{minipage}
\end{table*}

\negskip
\section{Conclusion}
\negskip

We reproduced BSARec for sequential recommendation, and systematically extended the analysis from the perspective of user history frequency, \ac{DSP} methods, and padding strategies.  
Overall, BSARec shows strong performance, with the frequency rescaler having positive effect on the architecture.
The wavelet transform offers marginal improvements over the Fourier transform, likely due to better locality. However, employing a DSP method instead of a simple residual connection provides limited improvement.
Notably, the padding used affects the signal representation within the frequency rescaler, and an appropriate padding type can enhance performance.
Some limitations remain; the hyperparameters were tuned for encoded sequence length \(\ell = 50\) and re-used due to limited resources.
Future work could explore DSP methods to better capture locality, potentially benefiting tasks such as session-based SR. 
Finally, the results for alternative padding strategies for SR models using DSP methods are promising, and further investigation is needed to extend these findings across a broader range of SR models.

\section{Acknowledgements}
The authors thank Prof.\ Maarten de Rijke for constructive comments that improved the writing and presentation of this paper.